\documentclass[superscriptaddress,twocolumn,showpacs,prl]{revtex4-1}
\usepackage{graphicx}
\usepackage{amsmath}
\usepackage{amssymb}

\usepackage{times}
\begin{document}

\title{Robust quantum enhanced phase estimation in a multimode interferometer}
\author{J. J. Cooper}
\affiliation{School of Physics and Astronomy, University of Leeds, Leeds LS2 9JT, United Kingdom}
\author{D. W. Hallwood}
\affiliation{Centre for Theoretical Chemistry and Physics and NZIAS, Massey University, Private Bag 102904, North Shore, Auckland 0745, New Zealand}
\author{J. A. Dunningham}
\affiliation{School of Physics and Astronomy, University of Leeds, Leeds LS2 9JT, United Kingdom}
\author{J. Brand}
\affiliation{Centre for Theoretical Chemistry and Physics and NZIAS, Massey University, Private Bag 102904, North Shore, Auckland 0745, New Zealand}
\pacs{03.75.Dg,06.20.Dk,03.65.Ta,03.75.Ss}

\begin{abstract}
By exploiting the correlation properties of ultracold atoms in a multi-mode interferometer, we show how quantum enhanced measurement precision can be achieved with strong robustness to particle loss. While the potential for enhanced measurement precision is limited for even moderate loss in two-mode schemes, multi-mode schemes can be more robust. A ring interferometer for sensing rotational motion with non-interacting fermionic atoms can realize 
an uncertainty scaling of $1/(N\sqrt{\eta})$ for $N$ particles with a fraction $\eta$ remaining after loss, which undercuts the shot noise limit of two mode interferometers. A second scheme with strongly-interacting bosons achieves a comparable measurement precision and improved readout.
\end{abstract}
\maketitle

Quantum interferometry aims at improving measurement precision with limited resources by way of quantum entanglement of the constituent particles, say photons or ultra-cold atoms \cite{Giovannetti2004, Holland1993, Dunningham2002}. However, entanglement is rapidly degraded by decoherence, 
e.g.\ due to particle loss. This makes it difficult to employ entanglement practically. Here we show how this limitation can be circumvented with robust multi-mode quantum superposition states. 

Interferometers estimate a parameter, e.g.  a phase angle $\phi$, by observing an interference pattern. A useful tool for quantifying the attainable quality of precision with a limited number of quanta $N$ is the quantum Fisher information (QFI), $F_Q$. This is the quantum analogue of the classical Fisher information and it is related to the uncertainty of the phase measurement via the Cram\'er-Rao bound, 
$\delta\phi \ge 1/\sqrt{F_Q}$~ \cite{Helstrom1976}. 
For unentangled particles (or independent single-quanta measurements) in the presence of loss,  $F_Q=N \eta$, where $\eta$ is the fraction of particles remaining. This yields the shot-noise limit $\delta\phi \ge 1/\sqrt{N \eta}$.

In two-mode interferometers, quantum-entangled input states  can improve this to a maximum QFI of $N^2$, called the Heisenberg limit~\cite{Yurke86,Lee2002, Giovannetti2004,Pezze09}. This provides the lowest uncertainty allowed by quantum mechanics for linear interferometry schemes, i.e.\ when $\phi$ is a single-particle observable \cite{Boixo2008}. The Heisenberg limit is achieved with maximally entangled states known as NOON states \cite{Lee2002, Giovannetti2004,Pezze09}. However, these are difficult to make~\cite{hallwood2007} and fragile to particle loss, reducing the QFI to 
$F_Q=N^2 \eta^N$~\cite{Dorner2009}. In the presence of only a small amount of loss ($\eta<1$) the NOON state provides less precision for large particle numbers than unentangled particles. Improving the robustness of entangled states is the subject of much recent work~\cite{Lee2009,frowis2011,cooper2010,Demkowicz2009,Dorner2009, Escher2011,Huelga1997}. However, even when optimizing the input state for a given loss and particle number~\cite{Demkowicz2009,Dorner2009, Escher2011}, the best achievable scaling is $F_Q \sim N$ for large $N$ \cite{Kolodynski2010}. Thus $F_Q \sim N^2$ is unattainable for two-mode schemes in the presence of loss. 

This limitation does not apply to multi-mode schemes~\cite{soderholm2003}.
If sensitivity scales with mode number and arbitrary modes are accessible, there are no strict limits on the QFI for either classical (unentangled) or quantum input states.
Unentangled states still suffer from a shot noise limit with the familiar scaling $\sim N$, when the particle number is varied in a given set of modes. Increasing the QFI can either be achieved by entanglement within the given set of modes or by accessing more sensitive modes, while the latter procedure is the more robust against losses. The key questions in this context is how to prepare input states where mode occupation scales with particle number. 
For fermions it is natural to occupy many modes due to Pauli correlations, which forbid the occupation of a single mode by more than one particle. For bosons, interactions lead to similar correlations and the occupation of many modes.

Here we demonstrate that a ring interferometer with non-interacting fermionic atoms can realize a QFI of $F_Q = N^2 \eta$, which retains sub-shot noise scaling 
of the uncertainty in the presence of particle loss. Furthermore, we show that using strongly interacting bosons produces a similar precision with improved readout resolution. 
Quantum correlations are essential to enhance the sensitivity beyond the shot noise limit of an unentangled (classical) input state by providing access to a larger mode space rather than maximizing entanglement.  

\emph{Ring interferometer -}
Specifically, let us consider a ring interferometer with ultra-cold atoms to measure rotation angles~\cite{ryu07}. Individual atoms, with an angular momentum $\hbar k$,  sense rotation by acquiring a phase shift $k\phi$. Here $\phi=\omega t_m$ is the rotation angle accumulated during the time interval $t_m$ and for a rotation rate $\omega$. A single-particle interferometry scheme could estimate $\phi$ through comparing, by interference, the phase shifts accumulated by different angular momentum modes, say $k_1$ and $k_2$. Since the phase shift difference is $(k_2-k_1)\phi$, we obtain a QFI of $F_Q=(k_2-k_1)N\eta$ in the case of $N\eta$ successfully repeated measurements. By adjusting the angular momentum difference $(k_2-k_1)\hbar$, the measurement precision can be amplified
without invoking entanglement, similar to Ref.~\cite{higgins2007}.
We are going to show below how the natural properties of correlated multi-particle systems
allow us to scale the amplification factor with the particle number in order to achieve 
a scaling beyond the shot noise limit, where $F_Q \propto N^2$. 
In contrast to NOON states, this scaling is not affected by loss.

We now consider a ring interferometer containing $N$ particles. 
For the purpose of interferometry, the system may be prepared in a binary superposition
$|\Psi\rangle = \frac{1}{\sqrt{2}}\left( |K_1\rangle + |K_2\rangle \right)$, where $|K\rangle=\sum_{\vec{n}}{\LARGE}^{(K)}C_{\vec{n}}|\vec{n}\rangle$ is the sum over the multi-index $\vec{n}=(...,n_{-1},n_{0},n_1,...)$ with fixed particle number, $\sum_k n_k = N$, and permanents $|\vec{n}\rangle = \prod_k (\hat{a}_k^{\dag})^{n_k}/\sqrt{n_k!}|\text{vac}\rangle$, where $\hat{a}_k^{\dag}$ creates a particle with angular momentum $\hbar k$. The 
notation $\sum^{(K)}$ implies the additional constraint $\sum_k n_k k = K$ fixing the total angular momentum. The precise composition of these states depends on the interactions and quantum statistics of the particles involved, and has important consequences for the robustness properties.

The superposition state is sensitive to rotation via the interaction term $\hat{H}_R = -\omega \hat{L}$, with the total angular momentum $\hat{L} = \sum_k \hbar k a^\dagger_k a_k$. This makes it suitable for rotation sensing~\cite{hallwood2009, cooper2010}. After the time interval $t_m$ it evolves into 
\begin{eqnarray}
|\Psi(\phi)\rangle = \frac{1}{\sqrt{2}}\left(e^{iK_1\phi} |K_1\rangle +e^{iK_2 \phi} |K_2\rangle \right),
\label{eq:bin_supPhase}
\end{eqnarray}
where an overall phase was ignored.

We can quantify the ability of the quantum state to precisely estimate the angle $\phi$ by the QFI. It is independent of the measurement procedure and is given by 
\begin{equation}
F_Q=4\left[ \langle\Psi'(\phi)|\Psi'(\phi)\rangle - \left|  \langle\Psi'(\phi)|\Psi(\phi)\rangle \right|^2 \right],
\label{FQpure}
\end{equation}
for a pure state, where $|\Psi'(\phi)\rangle = \partial|\Psi(\phi)\rangle/\partial \phi$~\cite{Braunstein1994}.

For the state given by Eq.~(\ref{eq:bin_supPhase}) we find 
$F_Q=(K_1-K_2)^2$, which gives $\delta \phi \ge 1/|K_1-K_2|$. To reach a scaling of $F_Q \sim N^2$ we, therefore, require $|K_1-K_2| = N$. For a two mode system, where the modes differ by only one unit of angular momentum, this is only possible with the NOON state.
This restraint
does not apply for states described by more than two modes, because $|K_1-K_2|=N$ can be achieved with many different configurations.  As will be shown, this allows huge improvements to the robustness of sub-shot noise limited measurements.

\emph{Particle Loss -} 
For a system coupled to a zero temperature environment the evolution of the system in the presence of particle loss is described by the master equation \cite{Barnett}
\begin{equation}
\dot{\rho} = \sum_k \frac{\Gamma}{2}\left[ 2\hat{a}_k\rho\hat{a}_k^\dag - \rho\hat{a}_k^\dag\hat{a}_k - \hat{a}_k^\dag\hat{a}_k\rho \right],
\label{master}
\end{equation}
where $\rho$ is the density matrix, $\dot{\rho}=\partial \rho/\partial t$, $\hat{a}_k$ is the annihilation operator of mode $k$, and $\Gamma$ is the loss rate, which is taken to be equal for all modes.  It was shown in Ref. \cite{Demkowicz2009} that it does not matter whether the loss occurs before, during or after the phase shift is acquired.  

Equation~(\ref{master}) is solved by describing $\rho$ by $N+1$ density matrices, $\rho^{(N-\nu)}$, having the particle number $N-\nu$,
\begin{equation}
\rho(t) = \sum_{\nu=0}^{N}g^{(N-\nu)}(t)\rho^{(N-\nu)}.
\end{equation}
Here $\rho^{(N-\nu)}=\frac{1}{N-\nu+1}\sum_{k=-\infty}^{\infty} \hat{a}_k \rho^{(N-\nu+1)} \hat{a}^\dag_k $ is time independent and normalized to $\text{Tr} \rho^{(N-\nu)} = 1$,
and $g^{(N-\nu)}(t)$ is a time dependent coefficient.  
Because the $\rho^{(N-\nu)}$ operate on distinct orthogonal subspaces, the
total QFI for a given loss rate is given by
\begin{equation}
F_{Q_{\eta}}=\sum_{\nu=0}^{N}g^{(N-\nu)}(t)F_Q^{N-\nu},
\label{FQeta}
\end{equation}
where $F_Q^{N-\nu}$ is the QFI of $\rho^{(N-\nu)}$ \cite{Demkowicz2009}.  From Eq.~(\ref{master}) it follows that
\begin{eqnarray}
\dot{g}^{(N-\nu)}(t) &=& \Gamma(N-\nu+1)g^{(N-\nu+1)}(t) \nonumber \\
&&-\Gamma(N-\nu)g^{(N-\nu)}(t),
\end{eqnarray}
which has the solution $g^{(N-\nu)}(t)=\binom{N}{\nu}\eta^{N-\nu}(1-\eta)^\nu$, where $\eta=e^{-\Gamma t}$. So $F_Q$ is easily determined for all $\eta$ if the $N+1$ values of $F_Q^{N-\nu}$ are known. We now compare the precision of unentangled atoms and NOON states with the precision of a fermionic and strongly-interacting bosonic superposition state in the presence of loss.

\emph{Fermionic System -}
Consider an  even number $N=2n$ of spin polarized fermions in the states $|-n\rangle = \prod_{i=-n}^{n-1}  \hat{a}_i^{\dag}|\text{vac}\rangle$ and $|n\rangle = \prod_{i=-n+1}^{n}  \hat{a}_i^{\dag}|\text{vac}\rangle$. 
Each one is a (shifted) Fermi sea with total angular momentum
$-n\hbar$ and $n\hbar$, respectively. The superposition $|\psi_{F} \rangle = \frac{1}{\sqrt{2}} (|-n\rangle+|n\rangle)$ thus has a QFI of $N^2$ and realizes sub-shot noise limited scaling. The state $|\psi_{F} \rangle $ can also be written as
\begin{align} \label{eq:psif}
|\psi_{F} \rangle &=\frac{1}{\sqrt{2}}\prod_{i=-n+1}^{n-1} \hat{a}_i^{\dag} \left( \hat{a}_{-n}^{\dag}+\hat{a}_{n}^{\dag} \right)|\text{vac}\rangle,
\end{align}
which shows that, effectively, only a single particle participates in the superposition. While the states $|-n\rangle$ and $|n\rangle$ are degenerate ground states of the kinetic energy $\hat{L}^2/(2mR^2)$ at zero rotation, the superposition  $|\psi_{F} \rangle$ emerges as the only ground state when the degeneracy is lifted in a weak external potential that breaks the rotational symmetry.

To evaluate the performance under loss, we consider the density matrix after removal of $\nu$ particles
\begin{eqnarray}
\rho^{(N-\nu)} &=& \frac{(N-\nu)!\nu!}{(N!)}\times \nonumber \\
&&\!\!\!\!\!\!\!\!\!\!\!\!\!\!\!\!\left[ \sum_{k_1<...<k_{\nu-1}=-n+1}^{n-1}  \hat{a}_{k_{\nu-1}}... \hat{a}_{k_{1}} |\chi \rangle \langle \chi|  \hat{a}_{k_1}^{\dag}... \hat{a}_{k_{\nu-1}}^{\dag}\right. \nonumber \\
&&\!\!\!\!\!\!\!\!\!\!\!+\frac{1}{2} \sum_{k_1<...<k_{\nu}=-n+1}^{n-1}  \left(  \hat{a}_{-n}^{\dag}+\hat{a}_n^{\dag} \right) \hat{a}_{k_{\nu-1}}... \hat{a}_{k_{1}} |\chi \rangle\otimes \nonumber \\
&&\left. \langle \chi|   \hat{a}_{k_1}^{\dag}... \hat{a}_{k_{\nu-1}}^{\dag}  \left(  \hat{a}_{-n}+\hat{a}_n \right)\right],
\end{eqnarray}
where $|\chi\rangle = \prod_{i=-n+1}^{n-1}\hat{a}_i^{\dag}|\text{vac}\rangle$ is a filled Fermi sea with $N-1$ particles. 

Since all terms are represented on mutually orthogonal subspaces, their QFI can be calculated individually and summed. We note that the first summation has no off-diagonal terms, so the QFI is simply zero. The second summation is a sum of $N-1 \choose \nu$ pure states, which all have QFI of $N^2$. This leads to $F_Q^{(N-\nu)} = \frac{N-\nu}{N} \times N^2$. Substituting into Eq.~\ref{FQeta} gives the QFI for a given loss,
\begin{eqnarray}
F_Q = N^2 \eta.
\end{eqnarray}
Significantly, the QFI is $N^2$ for no loss and decreases at the same rate as unentangled atoms when loss is considered. This is the expected result, because the state is a single-particle superposition with a difference in angular momentum of $N\hbar$. 
The role of the remaining atoms is to facilitate the creation of this superposition in the ground state.

Although this scheme gives an excellent precision and robustness to loss, the readout would require distinguishing the difference in momentum of a single particle, which is difficult. 
Furthermore, an even number of particles in the initial state needs to be ensured (e.g.\ by post-selection 
on the basis of total angular momentum) 
in order to exploit its degeneracy.
We now consider a scheme using strongly interacting bosonic atoms with improved readout options that is also insensitive to the even-odd particle number parity.

\begin{figure}
\centering
\includegraphics[scale=0.4]{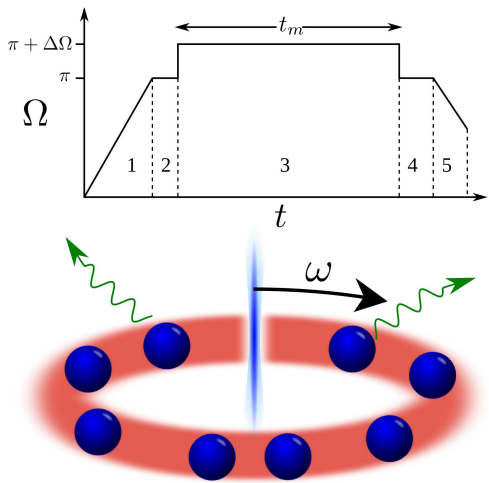}
\caption{Top:  A schematic showing the implementation of the scheme in time.  For a description of each step see the main text.  Bottom:  A visual representation of the system.  Ultra-cold atoms are confined to a 1D optical ring potential with a rotating barrier.  The measurement precision is determined in the presence of particle loss.}
\label{fig:schematic}
\end{figure}

\emph{Tonks-Girardeau System -} 
Bosonic atoms in a tightly confined wave guide at low densities and strong repulsive interactions realize the Tonks-Girardeau (TG) gas of impenetrable bosons~\cite{kinoshita2004}. The TG gas maps one-to-one to non-interacting spin polarized fermions \cite{girardeau_60}, and thus has similar properties. The non-rotating ground state, however, is non-degenerate and thus the previous scheme needs to be modified. A binary superposition with angular momentum difference of $N\hbar$ can be created with the help of a rotating narrow barrier potential as discussed in detail in Ref.~\cite{Hallwood2010}. Here, we calculate the QFI and suggest an  interferometry scheme that can take advantage of the theoretical precision bound.

Specifically, we consider $N$ bosonic atoms in a one-dimensional ring trap of circumference $L=2\pi R$
at zero temperature. They are stirred by a barrier of strength $b$, which rotates with angular velocity $\omega=h\Omega/(mL^2)$, where $\Omega$ is a phase induced around the ring in the co-rotating frame (see Fig.~\ref{fig:schematic}). The system is described by the Hamiltonian
\begin{eqnarray}
H &=& \sum_{k=-\infty}^{\infty} E_0 \left( k - \frac{\Omega}{2\pi} \right)^2 \hat{a}_k^\dag \hat{a}_k +  \frac{b}{L}\sum_{k_1,k_2=-\infty}^{\infty} \hat{a}_{k_1}^\dag \hat{a}_{k_2} \nonumber \\
&&+ \frac{g}{2L}\sum_{k_1,k_2,q=-\infty}^{\infty} \hat{a}_{k_1}^\dag \hat{a}_{k_2}^\dag \hat{a}_{k_1-q} \hat{a}_{k_2+q} ,
\label{eq:ham_loop}
\end{eqnarray}
where $E_0=2\pi^2\hbar^2/(mL^2)$ is the smallest non-zero kinetic energy of a single atom and $g>0$ is the interatomic interaction strength.

With external rotation at $\Omega = \pi$ and $b=0$, the ground state is degenerate between the states $|0\rangle$ with zero and $|N\rangle$ with $N\hbar$ angular momentum, regardless of the interaction strength. A finite barrier strength $0< b \ll g\sqrt{N}/2$ lifts the degeneracy and $|\psi_B^\pm\rangle =(|0\rangle\pm|N\rangle)/\sqrt{2}$ become eigenstates, where $|\psi_B^+\rangle$ is the ground state. This state has a QFI of $N^2$ and thus realizes sub-shot noise limited scaling. In the TG regime where formally $g\to \infty$, the Bose-Fermi mapping \cite{girardeau_60} allows us to map $|\psi_B^\pm\rangle$ onto a fermionic state similar to Eq.~(\ref{eq:psif}).

A measurement scheme for rotation is shown schematically in Fig.~\ref{fig:schematic}.  Step 1 corresponds to the creation of the initial state, 
where $\omega$ is slowly increased up to $\omega_0 = \pi h/(mL^2)$, corresponding to $\Omega = \pi$ (for details see Ref.~\cite{Hallwood2010}).
Once the initial state has been created, the state of the system will not change until the barrier's rotation rate is altered as indicated by step 2.  The additional rotation to be measured, $\Delta\omega$, is then non-adiabatically applied (step 3), inducing an additional phase of $\Delta\Omega$ around the ring.  The Hamiltonian describing the system is then 
$H'=\sum_{k=-\infty}^{\infty}E_0\left( k-\frac{1}{2}+\frac{\Delta\Omega}{2\pi} \right)^2\hat{a}_k^{\dag}\hat{a}_k$,
where we ignore the last two terms in Eq.~(\ref{eq:ham_loop}), because the barrier height is small and interactions do not couple states of different total angular momentum or change $|K\rangle$ when the rotation is changed. The system is then allowed to evolve 
for time $t_{m}$ as shown by step 3.  This establishes a phase difference between the two parts of the superposition and thus mixes $|\psi_B^+\rangle$  and $|\psi_B^-\rangle$.  At this point we calculate the QFI.

A possible readout scheme is to remove $\Delta\omega$ non-adiabatically leaving the barrier rotating at its original rate, $\omega_0$ (see step 4 in figure~\ref{fig:schematic}).  The rotation of the barrier is then adiabatically reduced to a point where the states $|\psi_B^+\rangle$  and $|\psi_B^-\rangle$ evolve into $|0\rangle$ and $|N\rangle$, respectively (step 5). The trapping potential is then removed and the atoms are imaged. The $|0\rangle$ state will have a peak at the center of the image while $|N\rangle$ will have a dip due to the different momentum distributions. By adiabatically reducing the interaction strength $g$ to a small value before removing the confining potential, the states $|0\rangle$ and $|N\rangle$ can be transformed into $(\hat{a}_0^\dag)^N|\text{vac}\rangle$ and $(\hat{a}_1^\dag)^N|\text{vac}\rangle$, respectively. This allows for an efficient distinction of the two outcomes.
The QFI can be shown to saturate the classical Fisher information for the total angular momentum.

The ring interferometer scheme can, with appropriate modifications, also measure rotational phases with a NOON state  $[(\hat{a}_0^\dag)^N +  (\hat{a}_1^\dag)^N]|\text{vac}\rangle$, which is obtained for small interactions, or with
a state of unentangled particles
$[(\hat{a}_0^\dag  +\hat{a}_1^\dag)^N]|\text{vac}\rangle$, obtained for  $b\gg g\sqrt{N}/2$.

\begin{figure}
\centering
\includegraphics[width=8cm]{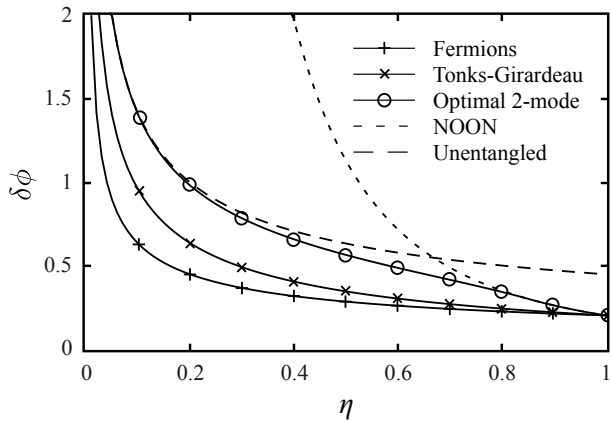}
\caption{The uncertainty of $\phi = \Delta \omega t_m$ for different fractions of particles remaining, $\eta$, and $N=5$.  The lines show $\delta\phi$ for non-interacting fermions and TG ($M=18$) superpositions compared with the optimal 2-mode (as given in \cite{Dorner2009}), NOON and unentangled states. The non-interacting fermions and TG ($M=18$) superpositions afford better precision than the unentangled 
particles  and the optimum two-mode initial state for all loss rates.}
\label{lossplot}
\end{figure}

Using Eq.~(\ref{FQeta}) we now compare the effects of particle loss (during time $t_m$) on the precision capabilities of the three states. 
The spread out momentum distribution of the TG state is different from fermions and limits numerical simulations of Eq.~(\ref{eq:ham_loop}) currently to 5 particles, where rescaling of $g$ ensures accurate results in a truncated basis of 18 momentum modes 
\cite{ernst11,Hallwood2010}. The QFI for a mixed state is
$F_Q=\text{Tr}[\rho(\phi)A^2]$, where $\rho(\phi)$ is the density matrix of the system and $A$ is the symmetric logarithmic negativity, defined as
$\frac{\partial \rho(\phi)}{\partial \phi} = \frac{1}{2}[A\rho(\phi) + \rho(\phi)A]$ \cite{Braunstein1994}.
In the eigenbasis of $\rho(\phi)$ this is $(A)_{ij} = 2[\rho'(\phi)]_{ij}/(\lambda_i+\lambda_j)$, where $\lambda_{i,j}$ are the eigenvalues of $\rho(\phi)$ and $\rho'(\phi) = \partial \rho(\phi)/\partial \phi$.  If $\lambda_i+\lambda_j=0$ then $(A)_{ij}=0$.

The results are shown in Fig.~\ref{lossplot} for $N=5$.  As expected the precision of the NOON, fermionic and TG superposition states are equivalent when there is no loss, $\eta=1$, and the precision of the unentangled state is much worse.  As the loss rate increases, however, the unentangled state soon outperforms the NOON state.  Importantly, the fermionic and TG superposition states outperform the unentangled state for all loss rates and therefore could prove extremely valuable for metrology.

For comparison with previous work, we also show the optimized initial two-mode state of Refs.~\cite{Dorner2009, Lee2009} for $N=5$  in Fig.~\ref{lossplot}.
We see that whilst it outperforms the NOON state, for stronger loss its precision gradually approaches that of an unentangled state as the fraction of particles remaining, $\eta$, decreases.  Importantly the precision of the fermionic and TG superposition states are better than the precision of the optimised two-mode initial state for all loss rates.  Not only this, the optimal two-mode initial state very much depends on the amount of loss as the structure of the state changes with $\eta$ thereby making its experimental implementation difficult.

\emph{Conclusion -} We have shown that robust sub-shot noise limited measurements 
are made possible by strong correlations of fermionic atoms due to the Pauli exclusion principle, and in TG systems due to strong interactions. The proposed states not only offer improved scaling of measurement precision, but also outperform optimized two mode states for small particle numbers. This is of significance for metrology as any state offering an increase in precision over unentangled states has the potential to alter the way precision measurements are made.  

This work was partly supported by EuroQUASAR and by the Marsden Fund (contract MAU0706), administered by the Royal Society of New Zealand.  JJC and JAD thank CTCP at Massey University for hospitality.




\begin{thebibliography}{20}

\bibitem{Giovannetti2004} V. Giovannetti, S. Lloyd and L. Maccone, Science \textbf{306}, 1330 (2004); Nature Photonics \textbf{5}, 222 (2011).

\bibitem{Holland1993} M.J. Holland and K. Burnett, Phys. Rev. Lett. \textbf{71}, 1355 (1993).

\bibitem{Dunningham2002} J.A. Dunningham, K. Burnett, and S.M. Barnett Phys. Rev. Lett. \textbf{89}, 150401 (2002).

\bibitem{Helstrom1976} C. W. Helstrom, Quantum Detection and Estimation Theory, Academic Press, New York (1976).

\bibitem{Yurke86} B. Yurke, Phys. Rev. Lett. \textbf{56}, 1515 (1986).

\bibitem{Lee2002} H. Lee, P. Kok and J. P. Dowling, J. Mod. Opt. \textbf{49}, 2325 (2002).

\bibitem{Pezze09} L. Pezz{\'e} and A. Smerzi, Phys. Rev. Lett. \textbf{102}, 100401 (2009).

\bibitem{Boixo2008} S. Boixo \textit{et al.}, Phys. Rev. A \textbf{77}, 012317 (2008). 

\bibitem{hallwood2007} D. W. Hallwood, K. Burnett and J. Dunningham, J. Mod. Opt. \textbf{54}, 2129 (2007). 


\bibitem{Dorner2009} U. Dorner \emph{et al.}, Phys. Rev. Lett. \textbf{102}, 040403 (2009).

\bibitem{Demkowicz2009} R. Demkowicz-Dobrza\'nski \emph{et al.}, Phys. Rev. A \textbf{80}, 013825 (2009).

\bibitem{Escher2011} B. M. Escher, R. L. de Matos Filho and L. Davidovich, Nature Physics \textbf{7}, 406 (2011).

\bibitem{Lee2009} T.-W Lee \textit{et al.}, Phys. Rev. A \textbf{80}, 063803 (2009).

\bibitem{frowis2011} F. Fr\"{o}wis and W. D\"{u}r, Phys. Rev. Lett. \textbf{106}, 110402 (2011).

\bibitem{cooper2010} J. J. Cooper, D. W. Hallwood, J. A. Dunningham, Phys. Rev. A \textbf{81}, 043624 (2010).

\bibitem{Huelga1997} S. F. Huelga \emph{et al.}, Phys. Rev. Lett. \textbf{79}, 3865 (1997).

\bibitem{Kolodynski2010} S. Knysh, V. N. Smelyanskiy, and G. A. Durkin, Phys. Rev. A \textbf{83}, 021804 (2011);
J. Ko\l{}ody\'{n}ski and R. Demkowicz-Dobrza\'{n}ski, {\it ibid.} \textbf{82}, 053804 (2010).

\bibitem{soderholm2003} J. S\"{o}derholm, G. Bj\"{o}rk, B. Hessmo, and S. Inoue, Phys. Rev. A \textbf{67}, 053803 (2003).

\bibitem{ryu07} C. Ryu \emph{et al.}, Phys. Rev. Lett. \textbf{99}, 260401 (2007).

\bibitem{higgins2007} B. L. Higgins, D. W. Berry, S. D. Bartlett, H. M. Wiseman and G. J. Pryde, Nature \textbf{450}, 393-396 (2007). 

\bibitem{hallwood2009} D. W. Hallwood, A. Stokes, J. J. Cooper and J. Dunningham, New J. Phys. \textbf{11}, 103040 (2009).

\bibitem{Braunstein1994} S. L. Braunstein, C. M. Caves and G. J. Milburn, Ann. Phys. (N.Y.) \textbf{247}, 135 (1996).

\bibitem{Barnett} S. M. Barnett and P. M. Radmore, \textit{Methods in Theoretical Quantum Optics}.

\bibitem{kinoshita2004} T. Kinoshita, T. Wenger, and D. S. Weiss, Science \textbf{305}, 1125 (2004). 

\bibitem{girardeau_60} M. Girardeau, J. Math. Phys. \textbf{1}, 516 (1960).

\bibitem{Hallwood2010} D. W. Hallwood, T. Ernst and J. Brand, Phys. Rev. A \textbf{82} 063623 (2010);
D. W. Hallwood and J. Brand, {\it ibid.}
\textbf{84}, 043620 (2011).

\bibitem{ernst11} T. Ernst, D. W. Hallwood, J. Gulliksen, H.-D. Meyer and J. Brand, Phys. Rev. A \textbf{84}, 023623 (2011).

\end{thebibliography}
\end{document}